\documentclass[aps,amsfonts,amsmath,prd,preprint,nofootinbib]{revtex4}
\pdfoutput=1
\newcommand{\beq}{\begin{equation}}
\newcommand{\eeq}{\end{equation}}
\usepackage{graphicx}
\usepackage{natbib}
\usepackage[utf8]{inputenc}
\begin{document}

\title{Spiky CMB distortions from primordial bubbles}
\author{Heling Deng}

\affiliation{Physics Department, Arizona State University, Tempe, AZ 85287, USA}

\begin{abstract}

Primordial bubbles that possibly nucleate through quantum tunneling during inflation in a multi-dimensional potential might have left some relic detectable at the present time. These bubbles turn into black holes during the radiation era, which may account for the LIGO black holes, supermassive black holes, and may play an important role in dark matter. Typically, these black holes are surrounded by an energy deficit in the form of a spherical sound wave packet propagating outwards. In this work we study how this perturbation of the cosmic plasma dissipates before the time of recombination, leading to spectral distortions in CMB. We find that there may exist some rare regions on the last scattering surface containing huge black holes, which have produced potentially detectable point-like signals of $\mu$-type distortions.

\end{abstract}

\maketitle

\section{Introduction}

Primordial black holes (PBHs) are hypothetical black holes formed in the early universe before large scale structures and galaxies.
Depending on the mechanism, a PBH could have a mass that ranges from the
Planck mass ($M_{\rm Pl}\sim10^{5}\ \text{g}$) to many orders of magnitude larger than the
solar mass $(M_{\odot}\sim10^{33}\ \text{g})$. PBHs with mass $M\lesssim10^{15}\ \text{g}$
would have evaporated by now due to Hawking radiation. The lack of
detected $\gamma$-rays from PBHs imposes stringent constraints on their
present density (see, e.g., ref. \cite{Carr:2020gox}), hence attempts to constrain PBHs have mainly been focused on nonevaporating ones.\footnote{Some recent works \cite{Boudaud:2018hqb, DeRocco:2019fjq, Laha:2019ssq, Dasgupta:2019cae} show that constraints from Galactic positrons could put slightly more stringent bounds near $M\sim 10^{15}\ \text{g}$ than the
extragalactic $\gamma$-ray background.}

Recent interest in PBHs was largely stimulated by LIGO, which detected
gravitational waves from inspiraling and merging black holes of masses
$\mathcal{O}(10 \mbox{--}100)\ M_{\odot}$ (e.g., \cite{Abbott:2016blz, Abbott:2016nmj, Abbott:2017vtc, Abbott:2017oio}). These masses are slightly larger than what
one would expect for ordinary stellar black holes, so LIGO black holes might have a primordial origin \cite{Bird:2016dcv, Sasaki:2016jop, Clesse:2016vqa, Carr:2016drx}.

Another motivation
to consider PBHs comes from supermassive black holes (SMBHs) at the
center of many galaxies \cite{LyndenBell:1969yx, Kormendy:1995er}. Their masses range from around $10^{6}\ M_{\odot}$ to $10^{10}\ M_{\odot}$,
and observations indicate that many of them are already in place at high
redshifts (see ref. \cite{Kormendy:2013dxa} for a review). Standard accretion models are unable
to explain such large masses due to insufficient cosmic time \cite{Haiman:2004ve}. One
is then led to consider other possibilities, and PBHs is a natural
candidate \cite{Rubin:2001yw, Bean:2002kx, Duechting:2004dk, Clesse:2015wea, Carr:2018rid} for they could have large masses by birth.

In addition, it has also long been suggested that dark matter observed in galaxies and clusters may be (partly) made up of PBHs. At the moment there are stringent constraints on the dark
matter fraction in PBHs within the mass range $10^{15}\ \text{g} \mbox{--}10^4\  M_{\odot}$ based on dynamical, microlensing and astrophysical effects (see, e.g., refs. \cite{Carr:2016drx, Inomata:2017vxo, Carr:2020gox} and references therein). This possibility has been excluded for most black hole masses, but some windows still remain.

PBHs can be formed in a variety of mechanisms. The most popular scenario is PBH formation during the radiation era from inflationary density perturbations \cite{Ivanov:1994pa, GarciaBellido:1996qt, Kawasaki:1997ju, Yokoyama:1998pt, Garcia-Bellido:2017mdw, Hertzberg:2017dkh}. After inflation, a large overdensity
($\delta \equiv \delta\rho/\rho\lesssim1$)  of superhorizonal scales can overcome pressure and collapse into
a black hole at horizon reentry. As is well known, density perturbations on scales $10^{-4}\ \text{Mpc}^{-1}\lesssim k \lesssim 1\ \text{Mpc}^{-1}$ are the seeds of large scale structures observed today, and the rms amplitude $\delta_{\text{rms}}$ is as small as $10^{-5}$. On the other hand, the formation of PBHs requires $\delta_{\text{rms}} > 0.1$ on small scales, assuming the primordial perturbations are Gaussian. One then has to construct inflationary potentials with this feature. However, as shown by refs. \cite{Kohri:2014lza, Nakama:2017xvq}, these models fail to account for SMBHs due to strong observational constraints on $\mu$-distortions in the CMB spectrum. This can be avoided if the primordial perturbations are non-Gaussian (e.g., \cite{Nakama:2016kfq}), or if PBHs are formed in other mechanisms after inflation (e.g., collapse of cosmic strings~\cite{Hawking:1987bn, Polnarev:1988dh, Garriga:1992nm, Caldwell:1995fu}, bubble collisions~\cite{Hawking:1982ga}, and collapse of closed domain walls \cite{Rubin:2000dq, Khlopov:2004sc}).

In this work, we will study the model developed in refs. \cite{Garriga:2015fdk,Deng:2016vzb,Deng:2017uwc}, in which PBHs are formed by primordial bubbles (spherical domain walls or vacuum bubbles) that nucleate during inflation via quantum tunneling. This model predicts distinctive PBH mass spectra ranging over many orders of magnitude, and can account for LIGO black holes, SMBHs and/or dark matter.\footnote{Ref. \cite{Kusenko:2020pcg} extended the work in ref. \cite{Deng:2017uwc} and constructed a PBH spectrum that is able to simultaneously explain all these three.}

Since quantum tunneling conserves total energy, each black hole should be compensated by an energy deficit. In this paper, we will focus on the scenario of domain wall bubbles\footnote{Much of the
analysis in this paper also applies to the scenario of vacuum bubbles. See section \ref{Conclusions and discussion}.} where each black holes is surrounded
by an underdense shell propagating outwards at the speed of sound. After time $t\sim6\times10^{6}\ \text{s}$, when the complete thermodynamic equilibrium between photons and the electromagnetic plasma can no longer be established, the dissipation
of the shell by photon diffusion injects heat into the plasma, potentially leading to detectable spectral distortions
in CMB. As we will see, this model satisfies the constraints imposed by observations, and predicts spiky distortions in some rare spots on the CMB sky, which is a unique feature of this model.\footnote{The anisotropy of spectral distortions are usually considered in the scenario of local-type primordial non-Gaussianity at small scales.} Estimating the magnitude of these distortions is the goal of the present paper.

We also note that besides being able to avoid the $\mu$-distortion constraint on primordial density perturbations, another attractive feature of our  model is that these PBHs have zero spin, which is not expected in most mechanisms proposed before. It was shown in ref. \cite{Pani:2013hpa} that the super-radiance of rotating black holes would release energy into the plasma in the early universe, resulting in spectral distortions. This enforces further constraints on the mass range $10^{-8} \mbox{--}0.2\ M_\odot$. Our model is apparently not subject to this bound.

The rest of the paper is organized as follows. The mechanism of PBH formation from spherical domain walls is reviewed in section \ref{The scenario and observational constraints}, along with the current observational constraints. CMB spectral distortions in this model are estimated in section \ref{Spectral distortions}. Conclusions are summarized and discussed in section \ref{Conclusions and discussion}. We set $c=\hbar=G=k_B=1$ throughout the paper.    

\section{The scenario and observational bounds \label{The scenario and observational constraints}}

Inflation is usually pictured as being driven by the evolution of a scalar field called the inflaton, which slowly rolls down its potential. The energy scale of the potential remains almost a constant during inflation, which causes a nearly exponential expansion. The field eventually ends up at a minimum of the potential corresponding to our vacuum. In general, the inflaton runs in a multi-dimensional potential, including a number of other minima. In a natural and simple scenario, the inflaton field is coupled to a double-well potential with two degenerate vacua. As the field slowly rolls down in one of the two wells along the direction that leads to our vacuum, it may tunnel to the other well, resulting in the nucleation of a spherical domain wall bubble \cite{Basu:1991ig}.


The formation of such a bubble can be thought of as a quantum process in a de Sitter background, where a spherical wall tunnels from a vanishing size to the Hubble size $H_i^{-1}$. It then gets stretched by the inflationary expansion. Bubbles formed at earlier times expand to larger sizes, and at the
end of inflation have a wide distribution \cite{Garriga:2015fdk, Deng:2016vzb},
\begin{equation}
n(R_{i})\approx\Gamma R_{i}^{-3}.
\end{equation}
where $n(R_{i})=R_{i}dn/dR_{i}$ is the number density of bubbles
of radius $\sim R_{i}$ when inflation ends, and $\Gamma$ is the
dimensionless bubble nucleation rate per Hubble volume per Hubble
time. We assume that $\Gamma$ is constant during inflation, which
is usually the case for small-field inflation.

Domain walls can be characterized by a finite energy density $\sigma$
per unit surface, and a tension of equal magnitude. In the thin wall limit,
the metric for a planar domain wall in the $yz$-plane (in an otherwise
empty space) is given by \cite{Vilenkin:1984hy,Ipser:1983db}
\begin{equation}
ds^{2}=-\left(1-H_{\sigma}\left|x\right|\right)^{2}dt^{2}+dx^{2}+\left(1-H_{\sigma}\left|x\right|\right)^{2}e^{2H_{\sigma}t}\left(dy^{2}+dz^{2}\right),
\end{equation}
where 
\begin{equation}
H_{\sigma}\equiv 2\pi \sigma.
\end{equation}
The $(x,t)$-part of the metric
is a $(1+1)$-dimensional Rindler space. A geodesic observer near the wall would be ``pushed away'' with acceleration $H_{\sigma}$. This means gravity produced by the wall is repulsive. In addition, we can see from the $(y,z)$-part of the metric that the hypersurface $x=0$, corresponding to the worldsheet of
the wall, expands exponentially at a constant rate $H_{\sigma}$.

When inflation ends, matter is created both inside and outside the wall due to the symmetry of the potential. Then the wall can be thought of as a comovingly static bubble living in an FRW universe dominated by radiation. Let  $t_{H}$ be the time when the bubble comes back within the Hubble horizon.
If $t_{H}\ll H_{\sigma}^{-1},$ the gravitational effect of the domain wall can be neglected. The bubble grows with the Hubble flow for a while, and at $t_{H}$ realizes it's actually a sphere rather than a planar wall. Then the surface
tension forces it to shrink to a black hole. Such bubbles are called
``subcritical''. On the other hand, a wall with $t_{H}\gg H_{\sigma}^{-1}$
is called ``supercritical'', and its gravitational effect becomes
significant at time $H_{\sigma}^{-1}$. Due to its repulsive nature,
the wall pushes the ambient radiation away, leaving two almost empty shells. The exterior FRW region continues its power law expansion, but the wall grows exponentially. This is possible only if a wormhole is created outside the domain wall bubble. The bubble grows without bound into a baby universe, and the wormhole eventually pinches off, turning into a black hole. 

In ref. \cite{Deng:2016vzb}, we found that the the black hole masses in this scenario can be approximated by
\begin{equation}
M\sim\begin{cases}
H_{\sigma}H_{i}^{2}R_{i}^{4}, & M_{\text{min}}<M<M_{*}\\
H_{i}R_{i}^{2}, & M>M_{*}
\end{cases},
\end{equation}
where $H_i$ is the Hubble expansion rate during inflation, $M_{*}\sim\mathcal{O}(10)H_{\sigma}^{-1}$ is the transition mass that connects the subcritical and supercritical regimes, and     $M_{\text{min}}\sim H_{\sigma}H_{i}^{-2}$ is the mass of the smallest black hole in this scenario (because the bubble forms with the Hubble size $H_i^{-1}$). Depending on the microphysics, the two characteristic masses $M_{*}$ and $M_{\text{min}}$ can have values within a wide range.

The mass distribution of black holes is conveniently characterized
by the mass function 
\begin{equation}
f(M)=\frac{M^{2}}{\rho_{{\rm {CDM}}}}\frac{dn(t)}{dM},\label{fdef}
\end{equation}
where $\rho_{{\rm {CDM}}}$ is the mass density of cold dark matter
(CDM). Here $M^{2}dn/dM$ can be interpreted as the mass density of
black holes in the mass range $\Delta M\sim M$. Since the black hole
density and $\rho_{{\rm {CDM}}}$ are diluted by the cosmic expansion
in the same way, $f(M)$ remains constant in time.

During the radiation era $(t<t_{{\rm eq}})$, the dark matter density
is of the order 
\begin{equation}
\rho_{{\rm {CDM}}}(t)\sim\left(Bt^{3/2}{\cal M}_{{\rm eq}}^{1/2}\right)^{-1},\label{rhoCDM}
\end{equation}
where $B\sim10$ is a constant and ${\cal M}_{{\rm eq}}\sim10^{17}\ M_{\odot}$
is the dark matter mass within a Hubble radius at $t_{{\rm {eq}}}$.
The mass function in our model is then given by
\begin{equation}
f(M)\sim B\Gamma{\cal M}_{{\rm eq}}^{1/2}\begin{cases}
M^{1/4}M_{*}^{-3/4}, & M_{\text{min}}<M<M_{*}\\
M^{-1/2}, & M>M_{*}
\end{cases}.\label{approxf-1}
\end{equation}

\begin{figure}[htb]
   \centering
   \includegraphics[scale=0.4]{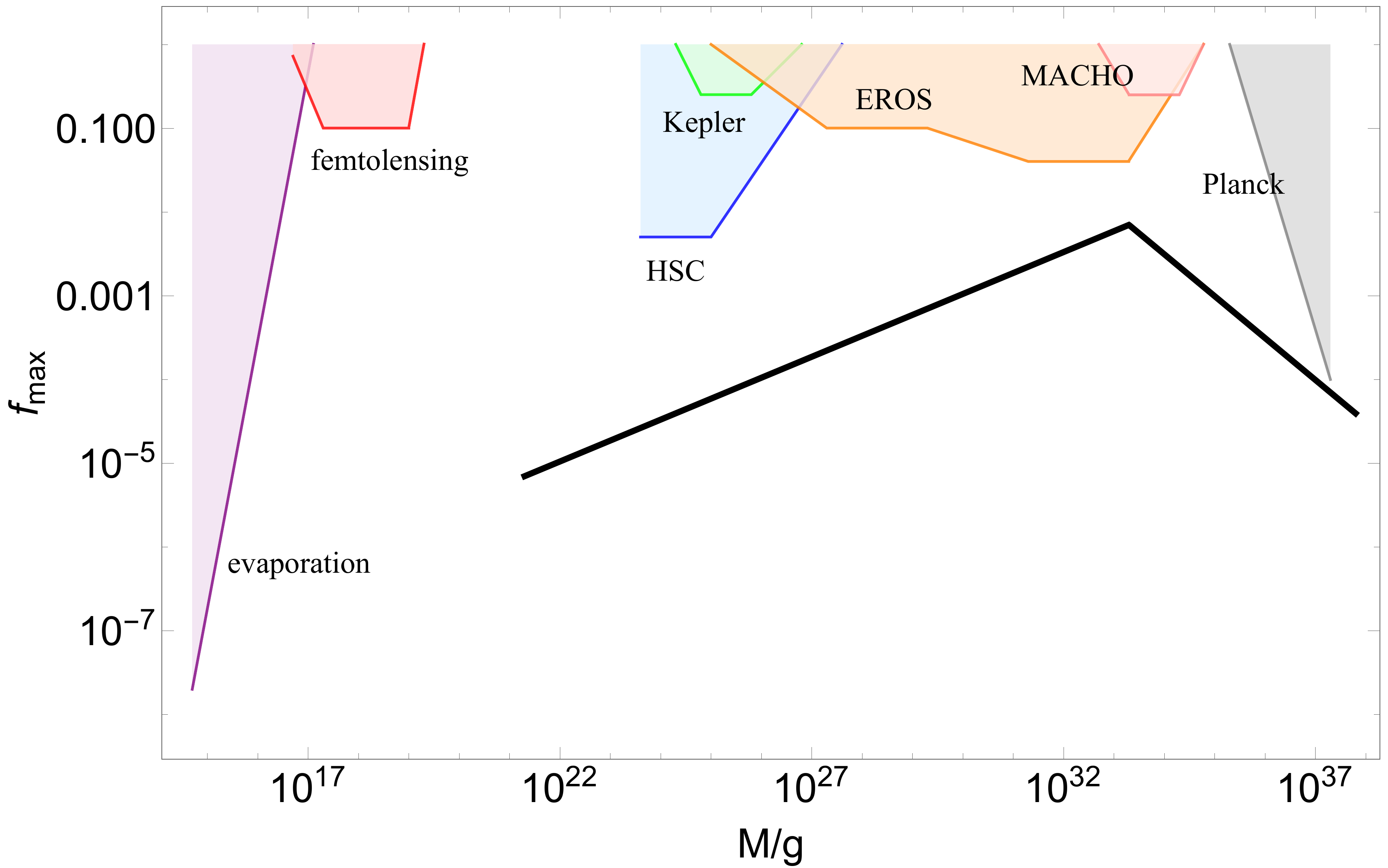}
   \caption{Constraints from different observations on the fraction of dark matter in PBHs as a function of the PBH mass for a monochromatic spectrum. More detail can be found in, e.g., ref. \cite{Carr:2020gox} and references therein. As an illustration, we also show the PBH mass function $f(M)$ for our domain wall model with $\Gamma=10^{-12}$, $M_{*}=M_{\odot}\approx10^{33}\text{ g}$ and $M_{{\rm min}}=10^{-12}M_{\odot}\approx10^{21}\text{ g}$.}
   \label{PBH2}
\end{figure}

Observational constraints on $f(M)$ within the mass range $10^{15}\ \text{g}\text{--}10^4\ M_\odot$ have been
extensively studied (see, e.g., ref. \cite{Carr:2020gox} for a recent review).
Some conservative bounds are summarized in fig. \ref{PBH2}. As an illustration, we
also show the distribution of the form (\ref{approxf-1}) with a set of parameters,
which can account for the LIGO observations (see below). Strictly
speaking, the bounds in fig. \ref{PBH2} only apply to a monochromatic
spectrum, where all black holes are of the same mass.
The method of using these bounds on an extended spectrum was investigated in ref. \cite{Carr:2017jsz} and
applied in, e.g., ref. \cite{Deng:2017uwc}, but the difference it brings is not significant.

LIGO black holes have masses around $30\ M_\odot$. In order to have the PBH merger rate suggested by the LIGO detectors, one requires \cite{Sasaki:2016jop}
\begin{equation}
f(M\sim30\ M_{\odot})\sim10^{-3},
\end{equation}
which can be consistent with the constraint from the Planck satellite (the rightmost bound in fig. \ref{PBH2}) \cite{Ricotti:2007au, Ali-Haimoud:2016mbv} only if
\begin{equation}
M_{*}\lesssim10^{2}\ M_{\odot},\ \ \ \Gamma\sim10^{-12}.
\end{equation}

The number density of PBH of mass $\sim M>M_{*}$ at the present time is
\begin{equation}
n_{M}^{(c)}\sim B\Gamma\left(\frac{\mathcal{M}_{\text{eq}}}{M}\right)^{1/2}\frac{\rho_{\text{CMB}}}{M}\sim10^{20}\Gamma\left(\frac{M}{M_{\odot}}\right)^{-3/2}\text{Mpc}^{-3}.\label{SMBH}
\end{equation}
The seeds for SMBHs should have density $n_{M}^{(c)}\sim0.1\ \text{Mpc}^{-3}$.
From eq. (\ref{SMBH}), the mass of such black holes in our scenario
is $M_{\text{seed}}\sim10^{14}\Gamma^{2/3}M_{\odot}$ (assuming that
$M_{\text{seed}}>M_{*}$). Requiring  $M_{{\rm seed}}\gtrsim10^{3}\ M_{\odot}$ \cite{Duechting:2004dk} yields $\Gamma\gtrsim10^{-17}$. The largest black hole we can expect
to find in the observable universe (of radius $\sim 10$~Gpc) has mass $M\sim10^{22}\Gamma^{2/3}M_{\odot}$. For $\Gamma \sim 10^{-12}$, this gives $M\sim 10^{14}\ M_\odot$.

Let $\eta_\sigma$ be the energy scale of the wall tension, and $\eta_i$ be the inflationary scale, then we have $H_i \sim \eta_i^2/M_{\rm Pl}$ and $H_\sigma \sim \eta_\sigma^3/M^2_{\rm Pl}$. An example of a set of parameters that may account for both  SMBHs and LIGO black holes is $\eta_\sigma \sim10^{7}$ GeV and 
$\Gamma\sim10^{-12}$. In this case, $M_{*}\sim M_{\odot}$, $f(M\sim30\ M_{\odot})\sim10^{-3}$,
and $M_{{\rm seed}}\sim10^{6}\ M_{\odot}$. The corresponding mass
distribution is shown in fig. \ref{PBH2}.  With these parameters, the black holes can also account for $\sim10\%$ of the dark matter.

Another interesting set of parameters is $\Gamma\sim10^{-17}$, $\eta_{i}\sim10^{7}$
GeV and $\eta_{\sigma}\sim10^{11}$ GeV, which give $M_{*}\sim M_{{\rm min}}\sim10^{19}$
g. In this case $f(M_{*})\sim1$, which means these black holes can constitute
all of the dark matter. Meanwhile, we have $M_{{\rm seed}}\sim10^{3}\ M_{\odot}$,
so they may also serve as seeds for SMBHs.

In the rest of the paper, we will show that, if the LIGO black holes were indeed formed by this mechanism, then as a consequence of the perturbation from the domain walls, there might exist some huge black holes on the boundary of the observable universe that are accompanied by detectable point-like distortions in the CMB spectrum.

\section{Spectral distortions from domain wall bubbles \label{Spectral distortions}}
CMB photons began their journey roughly $10^{13}\ \text{s}$ after the end of inflation, coming to us from the so-called last scattering surface (LSS) with comoving radius $\sim 10\ \text{Gpc}$ and thickness $\sim 20\ \text{Mpc}$. During the last scattering, photons decoupled from electrons, which combined with protons and formed atoms. Photons then travelled across the transparent universe in all directions, hardly interacting with other particles. Several generations of observations have revealed that the energy spectrum of the CMB is extremely close to a perfect black-body of temperature $T\approx 2.7$ K. Departures from the black-body spectrum, caused by the deviation from the equilibrium of photons and electrons, are commonly referred to as spectral distortions, which would encode important information from the early universe. Although no primordial distortions have been discovered, they could arise in plenty of processes. One important mechanism is the dissipation of sound waves. 

\begin{figure}[htb]
   \centering
   \includegraphics[scale=0.3]{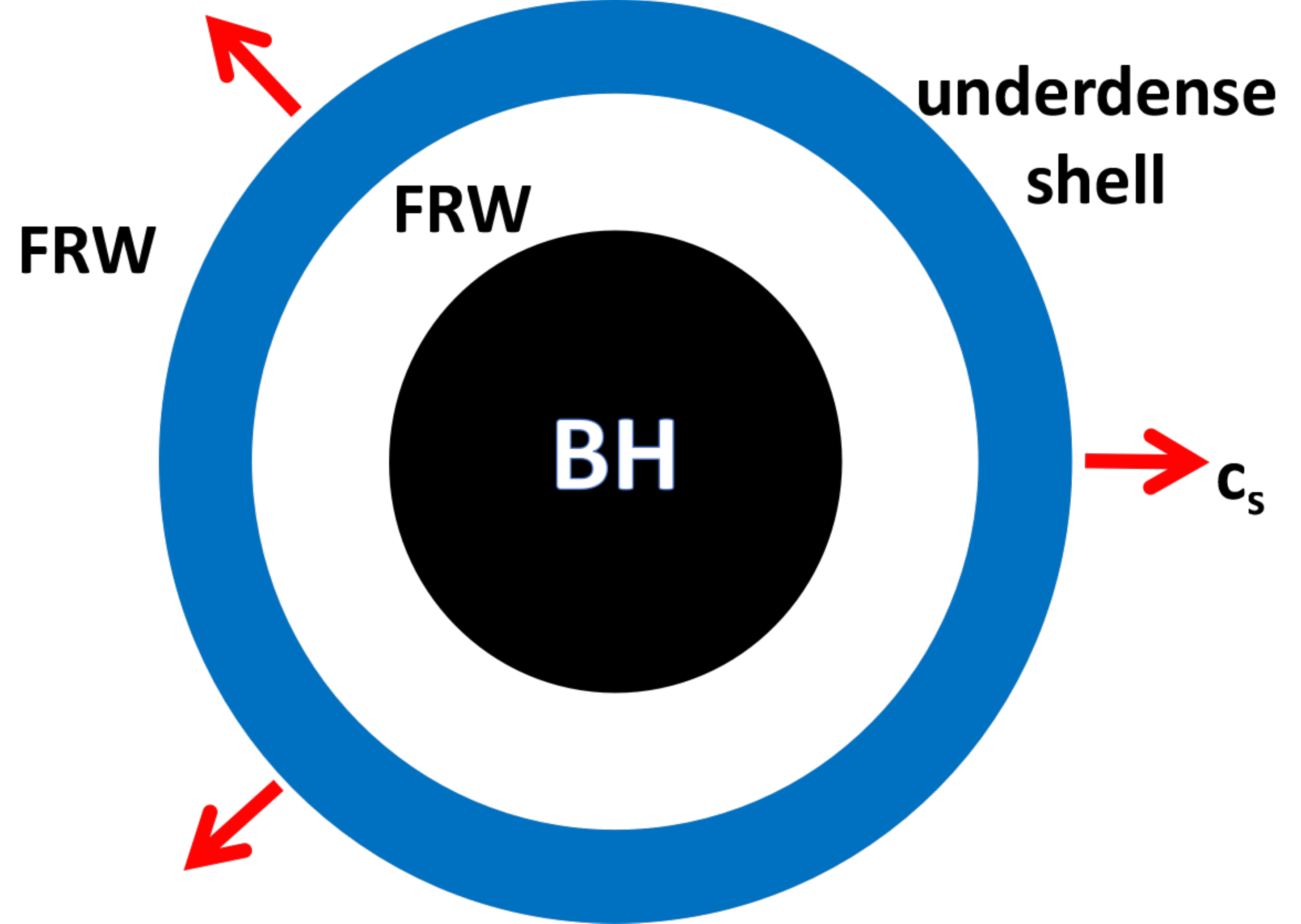}
   \caption{After inflation, a bubble turns into a black hole, and an underdense shell is produced as the bubble perturbs the ambient radiation. The energy deficit in the shell should roughly compensate the black hole mass. At late times the shell propagates outwards in the radiation dominated FRW (homogeneous) background as a spherical sound wave packet. }
   \label{spikybubble}
\end{figure}

PBH formation in our scenario stems from quantum tunneling, which conserves the global energy. For an outside observer not affected
by the domain wall, the central object has such a mass as if it is made
of normal FRW radiation. Roughly speaking, this implies that 
the resulting black hole should be compensated by some energy
deficit. In our setting, it was shown by our simulations that, when the black
hole is formed, it is surrounded by a shell of low density, and the
outer edge of the shell propagates outwards at the speed of sound $c_{s}=1/\sqrt{3}$  \cite{Deng:2016vzb}.
Simulations also showed that the black hole formation time $t_{M}$ is related to its mass $M$ by $t_{M}\sim3M$. Hence the wave front is at $\sim2c_{s}t_{M}\sim3M$ at $t_{M}$.
Considering that the black hole horizon is of size $\sim2M$, the thickness of the shell at $t_{M}$ is
\begin{equation}
s(t_{M})\sim M.
\end{equation}
As the wave front propagates outwards, radiation near the black hole
falls in and comes back to the FRW density. At late times, after it leaves the black hole, the shell
behaves like a spherical underdense sound wave packet (with a tiny
density contrast). Fig. \ref{spikybubble} shows a sketch of this picture. In an FRW universe, the shell's thickness gets redshifted by the cosmic expansion,
\begin{equation}
s(t)\sim\left(Mt\right)^{1/2}.
\end{equation}

Besides, it is well known that sound waves in a radiation dominated plasma
get dissipated by photon diffusion (also known as the Silk damping)
with a characteristic physical length scale 
\begin{equation}
\lambda(t)\sim\sqrt{\frac{t}{\sigma_{T}n_{e}}}\propto t^{5/4},\label{lambda}
\end{equation}
where $\sigma_{T}$ is the Thomson cross-section and $n_{e}$ is the
electron density. As the underdense shell propagates,
its thickness is smeared to a larger size. Effectively, the photon-electron
fluid can be approximately described by an imperfect fluid with a
shear viscosity \cite{Weinberg:1971mx}. The amplitude of the Fourier mode of a sound wave with physical wave number $k$ is approximately damped by $e^{-(k\lambda)^2}$, which is of the Gaussian form. It can be shown that, for a 1-D plane wave packet with a Gaussian profile (whose Fourier modes are also Gaussian), its thickness grows like  \cite{Deng:2018cxb}
\begin{equation}
d(t)=\sqrt{\lambda^{2}(t)+s^{2}(t)},
\end{equation}
and its amplitude gets damped by a factor $d(t)$. As a result, the area of the 1-D packet stays constant. In our setting, this implies the energy deficit of the shell is unchanged ($\sim M$).
The density contrast in the shell is then given by
\begin{equation}
\delta_{M}(t)\sim\frac{M}{\rho_{r}(t)\cdot4\pi(2c_{s}t)^{2}d(t)}=\frac{2M}{d(t)},
\end{equation}
where $4\pi(2c_{s}t)^{2}d(t)$ is the volume of the shell, and $\rho_{r}(t)=3/32\pi t^{2}$
is the background density. 

The sound wave energy density is given by $\rho_s=\rho_r \delta_M^2/4$, hence the total sound wave energy in a spherical shell can be estimated as
\begin{equation}
E_{Ms}(t) \sim \frac{1}{4} \rho_r(t) \delta_M^2(t) \cdot 4\pi(2c_{s}t)^{2}d(t) \sim \frac{M^2}{2d(t)},\label{sound_energy}
\end{equation} 
which is smaller than $M$ because $d(t)>M$.

Although the linearized energy deficit
of the underdense shell is a constant $(M\propto \delta_M)$, the sound wave energy ($E_{Ms}\propto \delta_M^2$) gets dissipated by photon diffusion and turns
into heat released into the background plasma, leading to spectral distortions. Depending
on the cosmic age when this happens, distortions have different
forms. These shall be the topics of the upcoming subsections.

\subsection{$\mu$-distortion} \label{mu-distortion}

$\mu$-distortion in CMB is produced by the mixture of photons of
different temperatures during the period known as the $\mu$-era, when $t_{\text{th}}\lesssim t\lesssim t_{\mu}$. At time $t<t_{\text{th}}\sim6\times10^{6}$
s $(\text{the corresponding redshift being }z>2\times10^{6})$ the
mixture is completely thermalized and we have a perfect black-body spectrum. At $t_{\text{th}}\lesssim t\lesssim t_{\mu}\sim10^{10}$ s $(5\times10^{4}<z<2\times10^{6})$, photon number changing processes
(like bremsstrahlung and double Compton scattering) become ineffective, while the Compton process still allows photons to reach equilibrium with the plasma, resulting in a spectrum with  nonzero chemical potential where there is energy release.

In our scenario, the comoving radius of the shell's wave front (which is also approximately the comoving sound horizon) $r(t)\sim 2c_{s}t/a(t)$ (where $a(t)$ is the scale factor) ranges from $r(t_{\text{th}})\sim0.1$~Mpc to $r(t_{\mu})\sim5$~Mpc during
the $\mu$-era; on the other hand, the photon diffusion scale  grows till the time of recombination, reaching the comoving value $\lambda_{\text{rec}}^{(c)}\equiv\lambda^{(c)}(t_{{\rm rec}})\sim 50$~Mpc.
(As a comparison, $\lambda^{(c)}(t_{{\rm th}})\sim0.55$~kpc and
$\lambda^{(c)}(t_{{\rm \mu}})\sim0.14$~Mpc \cite{Khatri:2015tla}.)  We shall refer to a spherical region of comoving radius $\lambda_{\text{rec}}^{(c)}$ as a Silk
region. Since $\lambda_{\text{rec}}^{(c)} \gg r(t_\mu)$, the damping of the shell acts effectively as a point-like energy release, and photons with
distorted spectra are eventually mixed within a Silk region containing black holes.  Let us now estimate the $\mu$-distortion from a single black hole.

Consider that a shell sweeps
through a comoving point at the wave front within the time
interval $(t,t+d/c_{s})$. The sound wave energy density is given by
\begin{equation}
\rho_{s}(t)=\frac{1}{4}\rho_{r}(t)\delta_{M}^{2}(t)=\rho_{r}\left[\frac{M}{d(t)}\right]^{2}.
\end{equation}
Then the energy density injected in the neighborhood of the point after the shell passes by is\footnote{Here we assume $d\ll t$, i.e., the cosmic expansion can be neglected
when the shell passes through the point. Later we will see the case
when this is not satisfied.}
\begin{equation}
\delta\rho_{s}(t)\sim \rho_{r}(t)M^{2}\left[d^{-2}(t)-d{}^{-2}\left(t+\frac{d}{c_{s}}\right)\right]\approx\frac{2\rho_{r}(t)M^{2}\dot{d}(t)}{c_{s}d^{2}(t)},\label{deltarho}
\end{equation}
where the overdot represents the first derivative with respect to time. So the dimensionless chemical potential $\mu$ ($\equiv-\mu_{\text{th}}/T$, where $\mu_{\text{th}}$ is the thermodynamic chemical potential) at this point is \cite{Sunyaev:1970er, Daly, Hu:1992dc, Chluba:2012we}
\begin{equation}
\mu(t)\approx1.4\frac{\delta\rho_{s}(t)}{\rho_{r}(t)}\approx\frac{2.8M^{2}\dot{d}(t)}{c_{s}d^{2}(t)}.\label{mu}
\end{equation}
The underdense wave packet propagates outwards, consecutively leaving
behind photons with different chemical potentials. These photons then
mix together. Consider a chemical potential $\mu$ produced at
comoving radius $r$.
Noting that the cosmic expansion doesn't change the value of $\mu$
at a certain point, the contribution from the comoving sphere with an infinitesimal thickness $dr$
to the photon mixture is $\propto\mu\cdot4\pi r^{2}dr$.
Therefore, the $\mu$-distortion averaged over a Silk region (with
comoving volume centered at
a single black hole of mass $\sim M$ is
\begin{equation}
\mu_{sM}=\frac{4\pi\int_{r_{\text{th}}}^{r_{\mu}}\mu r^{2}dr}{4\pi\lambda_{\text{rec}}^{(c)3}/3}=\frac{3\int_{t_{\text{th}}}^{t_{\mu}}\mu r^{2}\dot{r}dt}{\lambda_{\text{rec}}^{(c)3}},\label{musM0}
\end{equation}
where $r(t)=2c_{s}t/a(t)$, with $a(t)\propto t^{1/2}$. With the expression
of $\mu(t)$ in eq. (\ref{mu}), this gives
\begin{equation}
\mu_{sM}=\mathcal{O}(10)\left(\frac{M}{M_{\text{rec}}}\right)^{3/2}M^{1/2}\int_{t_{\text{th}}}^{t_{\mu}}\frac{\dot{d}(t)}{d^{2}(t)}t^{1/2}dt\equiv\mathcal{O}(10)\left(\frac{M}{M_{\text{rec}}}\right)^{3/2}I,\label{musM}
\end{equation}
where $M_{\text{rec}}\equiv\lambda^{2}(t_{\text{rec}})/t_{\text{rec}}\sim10^{17}\ M_{\odot}$,
and we have defined $I\equiv M^{1/2}\int_{t_{\text{th}}}^{t_{\mu}}\left[\dot{d}(t)t^{1/2}/d^{2}(t)\right]dt$.

\begin{figure}[htb]
   \centering
   \includegraphics[scale=0.3]{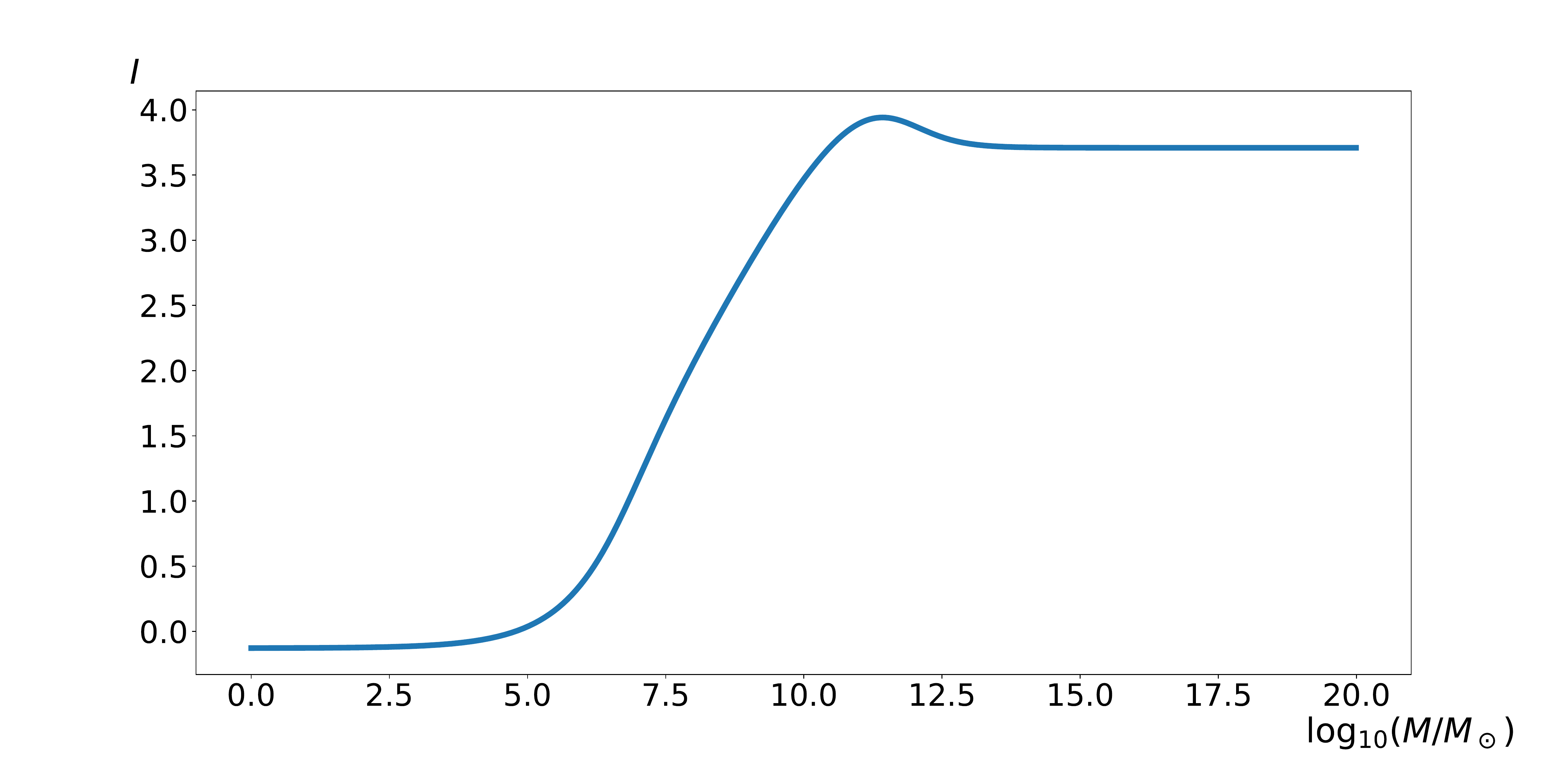}
   \caption{$I$ in eq. (\ref{integral}) as a function of the black hole mass $M$. Its value increases with $M$ till $M\sim 10^{12}\ M_\odot$ and approaches a constant.}
   \label{I}
\end{figure}

If there are many ($>1$) of these black holes within a Silk region,
the resulting $\mu$-distortion is the sum of their contributions.
In our model, the number density of supercritical black holes is given by
\begin{equation}
n_{M}(t)\sim\rho_{\text{CMB}}(t)\frac{f(M)}{M}\sim\Gamma\left(Mt\right)^{-3/2},
\end{equation}
then the total $\mu$-distortion from the contribution of black holes
of mass $\sim M$ is
\begin{equation}
\mu_{M}=\mu_{sM}\cdot n_{M}(t_{\text{rec}})\cdot\frac{4\pi}{3}\lambda^{3}(t_{\text{rec}})=\mathcal{O}(10)\Gamma I.\label{muM}
\end{equation}
Considering that $d(t)=\sqrt{\lambda^{2}(t)+s^{2}(t)},$ with $\lambda(t)\propto t^{5/4}$
and $s(t)=(Mt)^{1/2}$, the integral $I$ in eqs. (\ref{musM}) and
(\ref{muM}) can be found analytically, with the result
\begin{equation}
I=\left.\ln\left[t^{1/2}\left(\left(\frac{M_{t}}{M}+1\right)^{1/2}+1\right)^{-2/3}\right]-\left(\frac{M_{t}}{M}+1\right)^{-1/2}\right|_{t_{\text{th}}}^{t_{\mu}},\label{integral}
\end{equation}
where $M_{t}\equiv\lambda^{2}(t)/t$. We have $M_{\text{th}}\equiv \lambda^{2}(t_{\text{th}})/t_{\text{th}} \sim10^{7}\ M_{\odot}$, and $M_{\mu}\equiv \lambda^{2}(t_{\mu})/t_{\mu} \sim10^{12}\ M_{\odot}$. The function $I(M)$ is plotted in fig. \ref{I}. The physical meanings of $M_{\text{th}}$ and $M_\mu$ as well as the behavior of $I(M)$ will be
clear in the following two limits.

\subsubsection{$\lambda(t)\gg s(t)$ during $\mu$-era}

Although the photon diffusion length $\lambda\propto t^{5/4}$ is initially
small compared to the shell thickness $s\propto t^{1/2}$, it grows
faster and may at some point catch up with $s$. If this happens
before $t_{\text{th}}$, we have $d(t)\sim\lambda(t)$
during the $\mu$-era. Then since $\lambda(t_{\text{th}})\gtrsim s(t_{\text{th}})\sim\left(Mt_{\text{th}}\right)^{1/2},$
the black hole mass should satisfy $M\lesssim M_{\text{th}} \ll M_{\mu}$. In this limit, only the second term on the right hand side of eq. (\ref{integral}) is important, and we have
\begin{equation}
I\sim\left(\frac{M}{M_{\text{th}}}\right)^{1/2}.\label{I1}
\end{equation}
Therefore, by eq. (\ref{musM}), the $\mu$-distortion from a single
black hole is
\begin{equation}
\mu_{sM}\sim10^{-28}\left(\frac{M}{M_{\odot}}\right)^{2}.\label{musM1}
\end{equation}
If there are many of these black holes within a Silk region, then
by eq. (\ref{muM}), the total $\mu$-distortion is
\begin{equation}
\mu_{M}=\mathcal{O}(10)\Gamma\left(\frac{M}{M_{\text{th}}}\right)^{1/2}.\label{muM1}
\end{equation}
We can see that the largest $\mu_{M}$ is from the largest black hole in this limit.

\subsubsection{$\lambda(t)\ll s(t)$ during $\mu$-era}

On the other hand, if the photon diffusion length catches up with
the shell thickness after $t_\mu$, then we have $d(t)\sim s(t)$
during the $\mu$-era. Since $\lambda(t_{\mu})\lesssim s(t_{\mu}),$ the
black hole mass should satisfy $M\gtrsim M_{\mu}\gg M_{\text{th}}$.
In this limit, the right hand side of eq. (\ref{integral}) is dominated by $\ln t^{1/2}$, and we thus obtain
\begin{equation}
I\sim \frac{1}{2}\ln\left(\frac{t_{\mu}}{t_{\text{th}}}\right) \approx 3.\label{I2}
\end{equation}
This is independent of the black hole mass, so $I$ approaches a constant $\sim 3$ when $M\gtrsim 10^{12}\ M_\odot$. Then by eq. (\ref{musM}), the $\mu$-distortion from a single black
hole is
\begin{equation}
\mu_{sM}\sim10^{-24}\left(\frac{M}{M_{\odot}}\right)^{3/2}.\label{musM2}
\end{equation}


$\ $

The mass of the largest black hole in a Silk region ${\bar{M}}$ can
be found from eq.~(\ref{SMBH}) by setting $n_{M}^{(c)}\sim \left(4\pi \lambda_{\text{rec}}^{(c)3}/3\right)^{-1}$,
which gives ${\bar{M}}\sim 10^{17}\Gamma^{2/3}M_{\odot}$. In order
to account for the LIGO black holes in our model, we need $\Gamma\sim10^{-12}$,
which gives $\bar{M}\sim 10^{9}\ M_{\odot}$. By equalizing eqs. (\ref{I1})
and (\ref{I2}), one finds $M\sim10^{8}\ M_{\odot} < \bar{M}$. Therefore, a
rough estimate of the $\mu$-distortion in a typical Silk region (which is also an all-sky $\mu$) can
be given by  eq. (\ref{musM2}), which yields
\begin{equation}
\bar{\mu}\sim 10^{-24}\left(\frac{\bar{M}}{M_{\odot}}\right)^{3/2} \sim 10^{-11}.\label{mu_bar}
\end{equation}
This is much smaller than the value predicted from the Silk damping of sound waves in the standard $\Lambda$CDM
model ($\bar{\mu}\sim 10^{-8}$) \cite{Chluba:2016bvg}, and is thus too small to be detected.

However, a Silk region is but a tiny patch on the CMB sky. There will be regions that contain black holes with $M\gg\bar{M}$. Typically, the mass of the largest black hole $M_{\rm max}$ one can find on the LSS can be estimated by setting $n_M^{(c)} \sim \left[4 \pi \left(10\ {\rm Gpc}\right)^2 \times 50\ {\rm Mpc} \right]^{-1}$, where $10\ {\rm Gpc}$ is the radius of the LSS. With $\Gamma \sim 10^{-12}$, this gives $M_{\rm max}\sim 3\cdot 10^{12}\ M_{\odot}$,
which, by eq. (\ref{musM2}), gives the largest spiky $\mu$-distortion on the CMB sky,
\begin{equation}
\mu_{\text{max}}\sim10^{-24}\left(\frac{M_{\text{max}}}{M_{\odot}}\right)^{3/2}\sim10^{-6} \mbox{--}10^{-5}.\label{mu_max}
\end{equation}
Unlike $\bar{\mu}$, this is larger than $10^{-8}$ and hence potentially detectable.

A caveat of the above estimate is the following. For a mass as large
as $M_{\text{max}},$ the black hole happens to be formed
at $t_{M}\sim t_{\text{th}},$ that is, the beginning of the $\mu$-era.
With such a large black hole, the thickness of the underdense shell
is larger than the photon diffusion length during the $\mu$-era,
and is given by $(Mt)^{1/2}$, which at $t_{\text{th}}$ is about the Hubble size.
In the above analysis we assume that the shell sweeps through a comoving
point within a time much smaller than the Hubble time, but this is
no longer true when the shell's thickness is comparable to the Hubble
size, as is the case here. This means the estimate in the second limit
above should be checked more carefully. Now we show
that the result in eq. (\ref{musM2}) should be enhanced by one order of magnitude.

In fact, the shell ranges from $2c_{s}t-(Mt)^{1/2}$ (the inner edge
of the shell) to $2c_{s}t$ (the wave front) and it sweeps through
a point within the time interval $(t,t+\Delta t)$. We then have
\begin{equation}
\frac{2c_{s}t}{a(t)}=\frac{2c_{s}(t+\Delta t)-\sqrt{M(t+\Delta t)}}{a(t+\Delta t)},
\end{equation}
where the cosmic expansion has been taken into account. This gives
\begin{equation}
t+\Delta t\sim\left(M^{1/2}+t^{1/2}\right)^{2}.
\end{equation}
Therefore, instead of eq. (\ref{deltarho}), the sound wave energy
density injected into the point should be given by
\begin{equation}
\delta\rho_{s}\sim\left.\rho_{r}(t)\left[\frac{M}{(Mt)^{1/2}}\right]^{2}\right|_{t+\Delta t}^{t}=\frac{3M}{32\pi}\left[t^{-3}-(t+\Delta t)^{-3}\right].
\end{equation}
At $t\sim t_{\text{th}}$, $\Delta t \sim t_{\text{th}}$. 
As the shell passes through the point, the background density $\rho_{s}(t)$ decreases only by a
factor of order $\mathcal{O}(1 \mbox{--}10)$. Therefore the chemical potential
at that point can be approximated by
\begin{equation}
\mu(t)\sim\frac{\delta\rho_{s}}{\rho_{r}(t)}\sim\frac{M}{t}\left[1-\left(\sqrt{\frac{M}{t}}+1\right)^{-6}\right].\label{mu-1-1-1}
\end{equation}
Therefore, by eq. (\ref{musM0}), the $\mu$-distortion averaged over
a Silk region is
\begin{equation}
\mu_{sM}  =\frac{\mathcal{O}(10)}{M_{\text{rec}}^{3/2}}\int_{t_{\text{th}}}^{t_{\mu}}\mu(t) t^{1/2}dt \equiv\mathcal{O}(10)\left(\frac{M}{M_{\text{rec}}}\right)^{3/2}J, \label{musM0-1-1}
\end{equation}
where the integral $J$ is defined by considering the expression of $\mu(t)$ in eq. (\ref{mu-1-1-1}). Numerically evaluating it for $M\sim M_{\rm max}$ gives $J \approx 30$, which is $\sim 10$ times larger than the value of $I$ in eq. (\ref{I2}). Therefore the value of $\mu$ within such a Silk region is
\begin{equation}
\mu_{\text{max}}\sim10^{-5}  \mbox{--}10^{-4},
\end{equation}
which is an order of magnitude larger than the result in eq. (\ref{mu_max}).

The current  observational upper bound on the all-sky distortion ${\bar\mu}$ is $9\times 10^{-5}$,  from COBE/FIRAS \cite{Fixsen:1996nj}. Although ${\bar\mu}$ from the domain wall bubbles ($\sim 10^{-11}$) is smaller than the prediction from $\Lambda$CDM ($\sim 10^{-8}$), point-like signals $\mu_{\text{max}}\sim10^{-5}$ in rare Silk patches (whose angular size in CMB is $\sim 0.2^\circ$) marginally satisfies the current bound ($\sim 10^{-4}$) \cite{Khatri:2015tla}, and can possibly be detected in the near future. If the signal from $M_{\rm max}$ gets contaminated, black holes with, say, $M\sim M_\text{max}/3$ would give a weaker distortion $\mu_\text{max}\sim10^{-6}$.

\subsection{$y$-distortion}

After $t_{\mu}$, even the Compton process ceases to be extremely efficient, so that photons can no longer reach equilibrium with the heated plasma. As a result, the
photon spectrum gets Compton-distorted, similar to the Sunyaev-Zeldovich
effect observed in galaxy clusters, characterized by the
$y$-type spectrum with the magnitude of distortion given by the so-called
$y$-parameter.

In our setting, as in the case of $\mu$-distortion,
large black holes are accompanied with a $y$-distortion produced by the scattering of photons off the heated electrons as
the underdense shell dissipates. For the largest black hole on the LSS, its mass $M_{\rm max}\sim M_\mu$, which means the thickness of the shell is approximately the photon diffusion scale at $t>t_\mu$. This lies in the first limit discussed in subsection \ref{mu-distortion}. Similar to eq. (\ref{musM}), and by considering eq. (\ref{I1}), the $y$-parameter from a single black hole of mass $\sim M$ within a Silk region is given by
\begin{equation}
y_{sM}\sim\frac{\mathcal{O}(10)}{M_{\text{rec}}^{3/2}}\int_{t_{\mu}}^{t_{\text{rec}}}\mu t^{1/2}dt = \mathcal{O}(10)\left(\frac{M}{M_{\text{rec}}}\right)^{3/2}\left( \frac{M}{M_\mu} \right)^{1/2}.
\end{equation}
For $M\sim M_{\rm max}$, this gives $\mu_{\rm max} \sim 10^{-6}$. However, it was shown in ref. \cite{Hill:2015tqa} that the $\Lambda$CDM model predicts an all-sky $\bar{y}\sim 10^{-6}$ (only one order of magnitude below the current upper bound from COBE/FIRAS). Therefore, the $y$-distortion from the domain wall bubbles are probably too weak to be detected. 


\subsection{Temperature fluctuations}

We finally consider the temperature fluctuation in the underdense shell on the CMB sky.   For a black hole with mass $M\sim 10^{12}\ M_\odot$, the (physical) thickness of the shell at recombination is $\lambda(t_{\rm rec}) \sim 10^{-3} \lambda^{(c)}_{\rm rec} \sim 0.05\ \text{Mpc}$ (angular scale $\sim 0.2^\circ$). The temperature fluctuation in this region is 
\begin{equation}
\frac{\delta T}{T} \sim \frac{1}{4} \delta_M(t_{\rm rec}) \sim \frac{M}{\lambda(t_{\rm rec})}\sim 10^{-6}.
\label{dT/T}
\end{equation}
The size of a shell on the LSS at $ t_{\rm rec}$ corresponds to an angular scale of $\sim 1^\circ$. Since this is several times larger than the thickness of the LSS, the energy deficit in the shell could induce ring-like temperature patterns on the CMB sky, with diameter $\sim 1^\circ$ and thickness $\sim 0.2^\circ$. The spiky $\mu$-distortion can be found at the center of the ring. It is also possible that, since the shell itself is also thicker than the LSS, the temperature fluctuation would be a round spot instead of a ring, with an angular size $\lesssim 1^\circ$.

\section{Conclusions and discussion \label{Conclusions and discussion}}

In this paper we have estimated the spectral distortions in CMB as a relic of primordial bubbles (spherical domain walls, in particular) that nucleate during inflation through quantum tunneling. These bubbles turn into black holes after Hubble reentry during the radiation era. Since quantum tunneling conserves the global energy, each black hole comes with an energy deficit that compensates the black hole mass. The energy deficit propagates outwards in the form of a spherical, underdense sound wave packet, and the dissipation of the wave due to photon diffusion releases energy into the background plasma. This would lead to both $\mu$-type and $y$-type distortions in CMB. In order not to violate the constraints from the LIGO observations, the bubble nucleation rate during inflation has an upper bound $\Gamma \lesssim 10^{-12}$, which, along with the observational constraints on the dark matter fraction in PBHs, roughly fixes the mass spectrum in this model. Our calculations show that the average magnitude of distortions is much smaller than what was predicted in the standard $\Lambda$CDM mode and so is unlikely to be observed.

However, bubbles formed earlier during inflation would have larger sizes and thus turn into huge black holes. On the last scattering surface, there could be some rare ($\sim 1$) Silk patches that contain a black hole of mass $M\sim 10^{12}\ M_\odot$, leading to localized peaks of distortions with $\mu_\text{max} \sim 10^{-5}$. Such a spiky distribution of spectral distortions would be a unique signal of our PBH model, and is potentially detectable in the near future.

       
Similar conclusions can as well be drawn for the scenario of vacuum bubbles. Such bubbles can be formed when the tunneling takes the field to a well with an energy density smaller than the inflationary scale. One of the main differences is that vacuum bubbles tend to grow at the speed of light, and thus run into the ambient radiation fluid with a large Lorentz factor after inflation ends. In ref. \cite{Deng:2018cxb} we studied in detail the case of ``solid'' bubbles, which remains a vacuum inside, and completely reflect the fluid outside. It was shown by simulations that the black hole spectrum in this scenario is the same as that for domain wall bubbles in the supercritical regime. As a bubble runs into the radiation, a shock (overdense) wave would be formed due to energy transfer, followed by an underdense shell. Spectral distortions can be produced in the same manner, except that there's additional dissipation of an overdense thin shell with energy excess $\sim M$. In ref. \cite{Deng:2018cxb}, we argued that, for black holes with mass $M<M_{\rm th}$, the effect comes mainly from the underdense shell; but for larger black holes, distortions are dominated by the dissipation of the overdense region, because the sound wave energy in the underdense shell is relatively negligible. Considering that (i) the thickness of the overdense shell is smeared to the photon diffusion length before the $\mu$-era (corresponding to the first limit in subsection \ref{mu-distortion}), (ii) the PBH mass function for supercritical black holes is the same as that in eq. (\ref{approxf-1}), and (iii) the overdense shell gets dissipated in the same way as the underdense shell, we can use eqs. (\ref{musM1}) and (\ref{muM1}) to estimate the $\mu$-distortion from vacuum bubbles. By eq. (\ref{mu_bar}), we come to the same conclusion as in ref. \cite{Deng:2018cxb} that the average $\mu$-distortion is $\sim 10^{-11}$ and would be negligible.

However, for the largest black hole on the LSS, the resulting $\mu_\text{max}$ was overestimated in ref. \cite{Deng:2018cxb}, which has the result $\mu_\text{max}= \mathcal{O}(0.1 \mbox{--}1)$ for $M_{\rm max}\sim 10^{12}\ M_\odot$. This apparently violates the observational bound $\mu_\text{max}\lesssim 10^{-4}$. On the other hand, by eq. (\ref{musM1}) in this paper, we have $\mu_\text{max}\sim 10^{-3}$.\footnote{As mentioned in subsection \ref{mu-distortion}, if the largest possible signal gets contaminated, black holes with $M\sim M_\text{max}/3$ would give a distortion weaker by a factor of 10.} The difference comes from the fact that, in ref. \cite{Deng:2018cxb}, we used $k_D^{-1}$ to characterize the photon diffusion scale, where $k_D$ is the comoving wave number that enters the Silk damping term $e^{-(k/k_D)^2}$ for a fluctuation mode with wave number $k$; while in this paper, we use $\lambda^{(c)} = 2\pi/k_D$. For instance, the comoving photon diffusion scale at recombination $\lambda^{(c)}_{\rm rec}$ is taken to be $50$ Mpc here, but in ref. \cite{Deng:2018cxb} we used $10$ Mpc. A change this leads to is in the ratio $M_{\rm max}/\bar{M}\propto \left(\lambda^{(c)}_{\rm rec} \right)^{-4/3}$, which we used in the relation $\mu_{\rm max}=\bar{\mu}(M_{\rm max}/\bar{M})^2$ to find the distortion from $M_{\rm max}$ in ref. \cite{Deng:2018cxb} . Consequently, the change of $\lambda^{(c)}_{\rm rec}$ decreases $\mu_{\rm max}$ by a factor of $5^{8/3}=\mathcal{O}(100)$. 

Yet another scenario of PBH formation is from vacuum bubbles that interact with standard model particles only gravitationally after inflation. In this case, the radiation fluid is able to penetrate the wall and freely flows into the bubble interior. As a result, after inflation ends, the bubble can grow into a much larger size before coming to a stop with respect to the Hubble flow. We expect the energy deficit created afterwards outside the bubble would behave in a similar way, as an underdense sound wave packet, and hence a spiky spectral distortion is expected in some rare regions. However, in this scenario the PBH mass function might have a different form, which will be left as future research.

Lastly we note that, a recent work \cite{Serpico:2020ehh} studies the disk and spherical accretion of a halo around PBHs, strongly constraining the dark matter fraction in PBHs in the mass range around $1\text{--}10^4\ M_\odot$, reaching $f_{\rm PBH} < 3\times 10^{-9}$ at $\sim 10^4\ M_\odot$. This is much more stringent than the bound shown in fig. \ref{PBH2}. Although it may be less secure than constraints from dynamical effects \cite{Carr:2020gox}, if such a bound were to be taken into account, the following two possibilities in our model may be considered. On the one hand, we could impose an upper cut-off on the black hole mass, which can be achieved by relaxing the assumption that the bubble nucleation rate is a constant during inflation \cite{Deng:2018cxb}. In this case, our PBHs could account for LIGO black holes but not SMBHs, and the resulting CMB spectral distortions would be too weak to be detected. On the other hand, we could shift the transition mass $M_*$ to a value much larger than $10^4\ M_\odot$. Then our PBHs could still be the seeds for SMBHs at the galactic centers, and spiky signals of CMB distortions could potentially be observed.

\section*{Acknowledgements}
I am grateful to Alex Vilenkin and Masaki Yamada for useful comments and discussion. This work is supported by the U.S. Department of Energy, Office of High Energy Physics, under Award No. DE-SC0019470 at Arizona State University.

\bibliography{reference}

\begin{thebibliography}{61}
\expandafter\ifx\csname natexlab\endcsname\relax\def\natexlab#1{#1}\fi
\expandafter\ifx\csname bibnamefont\endcsname\relax
  \def\bibnamefont#1{#1}\fi
\expandafter\ifx\csname bibfnamefont\endcsname\relax
  \def\bibfnamefont#1{#1}\fi
\expandafter\ifx\csname citenamefont\endcsname\relax
  \def\citenamefont#1{#1}\fi
\expandafter\ifx\csname url\endcsname\relax
  \def\url#1{\texttt{#1}}\fi
\expandafter\ifx\csname urlprefix\endcsname\relax\def\urlprefix{URL }\fi
\providecommand{\bibinfo}[2]{#2}
\providecommand{\eprint}[2][]{\url{#2}}

\bibitem[{\citenamefont{Carr et~al.}(2020)\citenamefont{Carr, Kohri, Sendouda,
  and Yokoyama}}]{Carr:2020gox}
\bibinfo{author}{\bibfnamefont{B.}~\bibnamefont{Carr}},
  \bibinfo{author}{\bibfnamefont{K.}~\bibnamefont{Kohri}},
  \bibinfo{author}{\bibfnamefont{Y.}~\bibnamefont{Sendouda}}, \bibnamefont{and}
  \bibinfo{author}{\bibfnamefont{J.}~\bibnamefont{Yokoyama}}
  (\bibinfo{year}{2020}), \eprint{2002.12778}.

\bibitem[{\citenamefont{Boudaud and Cirelli}(2019)}]{Boudaud:2018hqb}
\bibinfo{author}{\bibfnamefont{M.}~\bibnamefont{Boudaud}} \bibnamefont{and}
  \bibinfo{author}{\bibfnamefont{M.}~\bibnamefont{Cirelli}},
  \bibinfo{journal}{Phys. Rev. Lett.} \textbf{\bibinfo{volume}{122}},
  \bibinfo{pages}{041104} (\bibinfo{year}{2019}), \eprint{1807.03075}.

\bibitem[{\citenamefont{DeRocco and Graham}(2019)}]{DeRocco:2019fjq}
\bibinfo{author}{\bibfnamefont{W.}~\bibnamefont{DeRocco}} \bibnamefont{and}
  \bibinfo{author}{\bibfnamefont{P.~W.} \bibnamefont{Graham}},
  \bibinfo{journal}{Phys. Rev. Lett.} \textbf{\bibinfo{volume}{123}},
  \bibinfo{pages}{251102} (\bibinfo{year}{2019}), \eprint{1906.07740}.

\bibitem[{\citenamefont{Laha}(2019)}]{Laha:2019ssq}
\bibinfo{author}{\bibfnamefont{R.}~\bibnamefont{Laha}}, \bibinfo{journal}{Phys.
  Rev. Lett.} \textbf{\bibinfo{volume}{123}}, \bibinfo{pages}{251101}
  (\bibinfo{year}{2019}), \eprint{1906.09994}.

\bibitem[{\citenamefont{Dasgupta et~al.}(2019)\citenamefont{Dasgupta, Laha, and
  Ray}}]{Dasgupta:2019cae}
\bibinfo{author}{\bibfnamefont{B.}~\bibnamefont{Dasgupta}},
  \bibinfo{author}{\bibfnamefont{R.}~\bibnamefont{Laha}}, \bibnamefont{and}
  \bibinfo{author}{\bibfnamefont{A.}~\bibnamefont{Ray}} (\bibinfo{year}{2019}),
  \eprint{1912.01014}.

\bibitem[{\citenamefont{Abbott et~al.}(2016{\natexlab{a}})}]{Abbott:2016blz}
\bibinfo{author}{\bibfnamefont{B.~P.} \bibnamefont{Abbott}}
  \bibnamefont{et~al.} (\bibinfo{collaboration}{Virgo, LIGO Scientific}),
  \bibinfo{journal}{Phys. Rev. Lett.} \textbf{\bibinfo{volume}{116}},
  \bibinfo{pages}{061102} (\bibinfo{year}{2016}{\natexlab{a}}),
  \eprint{1602.03837}.

\bibitem[{\citenamefont{Abbott et~al.}(2016{\natexlab{b}})}]{Abbott:2016nmj}
\bibinfo{author}{\bibfnamefont{B.~P.} \bibnamefont{Abbott}}
  \bibnamefont{et~al.} (\bibinfo{collaboration}{Virgo, LIGO Scientific}),
  \bibinfo{journal}{Phys. Rev. Lett.} \textbf{\bibinfo{volume}{116}},
  \bibinfo{pages}{241103} (\bibinfo{year}{2016}{\natexlab{b}}),
  \eprint{1606.04855}.

\bibitem[{\citenamefont{Abbott et~al.}(2017{\natexlab{a}})}]{Abbott:2017vtc}
\bibinfo{author}{\bibfnamefont{B.~P.} \bibnamefont{Abbott}}
  \bibnamefont{et~al.} (\bibinfo{collaboration}{VIRGO, LIGO Scientific}),
  \bibinfo{journal}{Phys. Rev. Lett.} \textbf{\bibinfo{volume}{118}},
  \bibinfo{pages}{221101} (\bibinfo{year}{2017}{\natexlab{a}}),
  \eprint{1706.01812}.

\bibitem[{\citenamefont{Abbott et~al.}(2017{\natexlab{b}})}]{Abbott:2017oio}
\bibinfo{author}{\bibfnamefont{B.~P.} \bibnamefont{Abbott}}
  \bibnamefont{et~al.} (\bibinfo{collaboration}{Virgo, LIGO Scientific}),
  \bibinfo{journal}{Phys. Rev. Lett.} \textbf{\bibinfo{volume}{119}},
  \bibinfo{pages}{141101} (\bibinfo{year}{2017}{\natexlab{b}}),
  \eprint{1709.09660}.

\bibitem[{\citenamefont{Bird et~al.}(2016)\citenamefont{Bird, Cholis,
  Mu{\~n}oz, Ali-Ha{\"\i}moud, Kamionkowski, Kovetz, Raccanelli, and
  Riess}}]{Bird:2016dcv}
\bibinfo{author}{\bibfnamefont{S.}~\bibnamefont{Bird}},
  \bibinfo{author}{\bibfnamefont{I.}~\bibnamefont{Cholis}},
  \bibinfo{author}{\bibfnamefont{J.~B.} \bibnamefont{Mu{\~n}oz}},
  \bibinfo{author}{\bibfnamefont{Y.}~\bibnamefont{Ali-Ha{\"\i}moud}},
  \bibinfo{author}{\bibfnamefont{M.}~\bibnamefont{Kamionkowski}},
  \bibinfo{author}{\bibfnamefont{E.~D.} \bibnamefont{Kovetz}},
  \bibinfo{author}{\bibfnamefont{A.}~\bibnamefont{Raccanelli}},
  \bibnamefont{and} \bibinfo{author}{\bibfnamefont{A.~G.} \bibnamefont{Riess}},
  \bibinfo{journal}{Phys. Rev. Lett.} \textbf{\bibinfo{volume}{116}},
  \bibinfo{pages}{201301} (\bibinfo{year}{2016}), \eprint{1603.00464}.

\bibitem[{\citenamefont{Sasaki et~al.}(2016)\citenamefont{Sasaki, Suyama,
  Tanaka, and Yokoyama}}]{Sasaki:2016jop}
\bibinfo{author}{\bibfnamefont{M.}~\bibnamefont{Sasaki}},
  \bibinfo{author}{\bibfnamefont{T.}~\bibnamefont{Suyama}},
  \bibinfo{author}{\bibfnamefont{T.}~\bibnamefont{Tanaka}}, \bibnamefont{and}
  \bibinfo{author}{\bibfnamefont{S.}~\bibnamefont{Yokoyama}},
  \bibinfo{journal}{Phys. Rev. Lett.} \textbf{\bibinfo{volume}{117}},
  \bibinfo{pages}{061101} (\bibinfo{year}{2016}), \eprint{1603.08338}.

\bibitem[{\citenamefont{Clesse and Garc{\'\i}a-Bellido}(2017)}]{Clesse:2016vqa}
\bibinfo{author}{\bibfnamefont{S.}~\bibnamefont{Clesse}} \bibnamefont{and}
  \bibinfo{author}{\bibfnamefont{J.}~\bibnamefont{Garc{\'\i}a-Bellido}},
  \bibinfo{journal}{Phys. Dark Univ.} \textbf{\bibinfo{volume}{15}},
  \bibinfo{pages}{142} (\bibinfo{year}{2017}), \eprint{1603.05234}.

\bibitem[{\citenamefont{Carr et~al.}(2016)\citenamefont{Carr, Kuhnel, and
  Sandstad}}]{Carr:2016drx}
\bibinfo{author}{\bibfnamefont{B.}~\bibnamefont{Carr}},
  \bibinfo{author}{\bibfnamefont{F.}~\bibnamefont{Kuhnel}}, \bibnamefont{and}
  \bibinfo{author}{\bibfnamefont{M.}~\bibnamefont{Sandstad}},
  \bibinfo{journal}{Phys. Rev.} \textbf{\bibinfo{volume}{D94}},
  \bibinfo{pages}{083504} (\bibinfo{year}{2016}), \eprint{1607.06077}.

\bibitem[{\citenamefont{Lynden-Bell}(1969)}]{LyndenBell:1969yx}
\bibinfo{author}{\bibfnamefont{D.}~\bibnamefont{Lynden-Bell}},
  \bibinfo{journal}{Nature} \textbf{\bibinfo{volume}{223}},
  \bibinfo{pages}{690} (\bibinfo{year}{1969}).

\bibitem[{\citenamefont{Kormendy and Richstone}(1995)}]{Kormendy:1995er}
\bibinfo{author}{\bibfnamefont{J.}~\bibnamefont{Kormendy}} \bibnamefont{and}
  \bibinfo{author}{\bibfnamefont{D.}~\bibnamefont{Richstone}},
  \bibinfo{journal}{Ann. Rev. Astron. Astrophys.}
  \textbf{\bibinfo{volume}{33}}, \bibinfo{pages}{581} (\bibinfo{year}{1995}).

\bibitem[{\citenamefont{Kormendy and Ho}(2013)}]{Kormendy:2013dxa}
\bibinfo{author}{\bibfnamefont{J.}~\bibnamefont{Kormendy}} \bibnamefont{and}
  \bibinfo{author}{\bibfnamefont{L.~C.} \bibnamefont{Ho}},
  \bibinfo{journal}{Ann. Rev. Astron. Astrophys.}
  \textbf{\bibinfo{volume}{51}}, \bibinfo{pages}{511} (\bibinfo{year}{2013}),
  \eprint{1304.7762}.

\bibitem[{\citenamefont{Haiman}(2004)}]{Haiman:2004ve}
\bibinfo{author}{\bibfnamefont{Z.}~\bibnamefont{Haiman}},
  \bibinfo{journal}{Astrophys. J.} \textbf{\bibinfo{volume}{613}},
  \bibinfo{pages}{36} (\bibinfo{year}{2004}), \eprint{astro-ph/0404196}.

\bibitem[{\citenamefont{Rubin et~al.}(2001)\citenamefont{Rubin, Sakharov, and
  Khlopov}}]{Rubin:2001yw}
\bibinfo{author}{\bibfnamefont{S.~G.} \bibnamefont{Rubin}},
  \bibinfo{author}{\bibfnamefont{A.~S.} \bibnamefont{Sakharov}},
  \bibnamefont{and} \bibinfo{author}{\bibfnamefont{M.~{\relax Yu}.}
  \bibnamefont{Khlopov}}, \bibinfo{journal}{J. Exp. Theor. Phys.}
  \textbf{\bibinfo{volume}{91}}, \bibinfo{pages}{921} (\bibinfo{year}{2001}),
  \bibinfo{note}{[J. Exp. Theor. Phys.92,921(2001)]}, \eprint{hep-ph/0106187}.

\bibitem[{\citenamefont{Bean and Magueijo}(2002)}]{Bean:2002kx}
\bibinfo{author}{\bibfnamefont{R.}~\bibnamefont{Bean}} \bibnamefont{and}
  \bibinfo{author}{\bibfnamefont{J.}~\bibnamefont{Magueijo}},
  \bibinfo{journal}{Phys. Rev.} \textbf{\bibinfo{volume}{D66}},
  \bibinfo{pages}{063505} (\bibinfo{year}{2002}), \eprint{astro-ph/0204486}.

\bibitem[{\citenamefont{Duechting}(2004)}]{Duechting:2004dk}
\bibinfo{author}{\bibfnamefont{N.}~\bibnamefont{Duechting}},
  \bibinfo{journal}{Phys. Rev.} \textbf{\bibinfo{volume}{D70}},
  \bibinfo{pages}{064015} (\bibinfo{year}{2004}), \eprint{astro-ph/0406260}.

\bibitem[{\citenamefont{Clesse and Garc{\'\i}a-Bellido}(2015)}]{Clesse:2015wea}
\bibinfo{author}{\bibfnamefont{S.}~\bibnamefont{Clesse}} \bibnamefont{and}
  \bibinfo{author}{\bibfnamefont{J.}~\bibnamefont{Garc{\'\i}a-Bellido}},
  \bibinfo{journal}{Phys. Rev.} \textbf{\bibinfo{volume}{D92}},
  \bibinfo{pages}{023524} (\bibinfo{year}{2015}), \eprint{1501.07565}.

\bibitem[{\citenamefont{Carr and Silk}(2018)}]{Carr:2018rid}
\bibinfo{author}{\bibfnamefont{B.}~\bibnamefont{Carr}} \bibnamefont{and}
  \bibinfo{author}{\bibfnamefont{J.}~\bibnamefont{Silk}},
  \bibinfo{journal}{Mon. Not. Roy. Astron. Soc.}
  \textbf{\bibinfo{volume}{478}}, \bibinfo{pages}{3756} (\bibinfo{year}{2018}),
  \eprint{1801.00672}.

\bibitem[{\citenamefont{Inomata et~al.}(2018)\citenamefont{Inomata, Kawasaki,
  Mukaida, and Yanagida}}]{Inomata:2017vxo}
\bibinfo{author}{\bibfnamefont{K.}~\bibnamefont{Inomata}},
  \bibinfo{author}{\bibfnamefont{M.}~\bibnamefont{Kawasaki}},
  \bibinfo{author}{\bibfnamefont{K.}~\bibnamefont{Mukaida}}, \bibnamefont{and}
  \bibinfo{author}{\bibfnamefont{T.~T.} \bibnamefont{Yanagida}},
  \bibinfo{journal}{Phys. Rev.} \textbf{\bibinfo{volume}{D97}},
  \bibinfo{pages}{043514} (\bibinfo{year}{2018}), \eprint{1711.06129}.

\bibitem[{\citenamefont{Ivanov et~al.}(1994)\citenamefont{Ivanov, Naselsky, and
  Novikov}}]{Ivanov:1994pa}
\bibinfo{author}{\bibfnamefont{P.}~\bibnamefont{Ivanov}},
  \bibinfo{author}{\bibfnamefont{P.}~\bibnamefont{Naselsky}}, \bibnamefont{and}
  \bibinfo{author}{\bibfnamefont{I.}~\bibnamefont{Novikov}},
  \bibinfo{journal}{Phys. Rev.} \textbf{\bibinfo{volume}{D50}},
  \bibinfo{pages}{7173} (\bibinfo{year}{1994}).

\bibitem[{\citenamefont{Garcia-Bellido
  et~al.}(1996)\citenamefont{Garcia-Bellido, Linde, and
  Wands}}]{GarciaBellido:1996qt}
\bibinfo{author}{\bibfnamefont{J.}~\bibnamefont{Garcia-Bellido}},
  \bibinfo{author}{\bibfnamefont{A.~D.} \bibnamefont{Linde}}, \bibnamefont{and}
  \bibinfo{author}{\bibfnamefont{D.}~\bibnamefont{Wands}},
  \bibinfo{journal}{Phys. Rev.} \textbf{\bibinfo{volume}{D54}},
  \bibinfo{pages}{6040} (\bibinfo{year}{1996}), \eprint{astro-ph/9605094}.

\bibitem[{\citenamefont{Kawasaki et~al.}(1998)\citenamefont{Kawasaki, Sugiyama,
  and Yanagida}}]{Kawasaki:1997ju}
\bibinfo{author}{\bibfnamefont{M.}~\bibnamefont{Kawasaki}},
  \bibinfo{author}{\bibfnamefont{N.}~\bibnamefont{Sugiyama}}, \bibnamefont{and}
  \bibinfo{author}{\bibfnamefont{T.}~\bibnamefont{Yanagida}},
  \bibinfo{journal}{Phys. Rev.} \textbf{\bibinfo{volume}{D57}},
  \bibinfo{pages}{6050} (\bibinfo{year}{1998}), \eprint{hep-ph/9710259}.

\bibitem[{\citenamefont{Yokoyama}(1998)}]{Yokoyama:1998pt}
\bibinfo{author}{\bibfnamefont{J.}~\bibnamefont{Yokoyama}},
  \bibinfo{journal}{Phys. Rev.} \textbf{\bibinfo{volume}{D58}},
  \bibinfo{pages}{083510} (\bibinfo{year}{1998}), \eprint{astro-ph/9802357}.

\bibitem[{\citenamefont{Garcia-Bellido and
  Ruiz~Morales}(2017)}]{Garcia-Bellido:2017mdw}
\bibinfo{author}{\bibfnamefont{J.}~\bibnamefont{Garcia-Bellido}}
  \bibnamefont{and}
  \bibinfo{author}{\bibfnamefont{E.}~\bibnamefont{Ruiz~Morales}},
  \bibinfo{journal}{Phys. Dark Univ.} \textbf{\bibinfo{volume}{18}},
  \bibinfo{pages}{47} (\bibinfo{year}{2017}), \eprint{1702.03901}.

\bibitem[{\citenamefont{Hertzberg and Yamada}(2018)}]{Hertzberg:2017dkh}
\bibinfo{author}{\bibfnamefont{M.~P.} \bibnamefont{Hertzberg}}
  \bibnamefont{and} \bibinfo{author}{\bibfnamefont{M.}~\bibnamefont{Yamada}},
  \bibinfo{journal}{Phys. Rev.} \textbf{\bibinfo{volume}{D97}},
  \bibinfo{pages}{083509} (\bibinfo{year}{2018}), \eprint{1712.09750}.

\bibitem[{\citenamefont{Kohri et~al.}(2014)\citenamefont{Kohri, Nakama, and
  Suyama}}]{Kohri:2014lza}
\bibinfo{author}{\bibfnamefont{K.}~\bibnamefont{Kohri}},
  \bibinfo{author}{\bibfnamefont{T.}~\bibnamefont{Nakama}}, \bibnamefont{and}
  \bibinfo{author}{\bibfnamefont{T.}~\bibnamefont{Suyama}},
  \bibinfo{journal}{Phys. Rev.} \textbf{\bibinfo{volume}{D90}},
  \bibinfo{pages}{083514} (\bibinfo{year}{2014}), \eprint{1405.5999}.

\bibitem[{\citenamefont{Nakama et~al.}(2018)\citenamefont{Nakama, Carr, and
  Silk}}]{Nakama:2017xvq}
\bibinfo{author}{\bibfnamefont{T.}~\bibnamefont{Nakama}},
  \bibinfo{author}{\bibfnamefont{B.}~\bibnamefont{Carr}}, \bibnamefont{and}
  \bibinfo{author}{\bibfnamefont{J.}~\bibnamefont{Silk}},
  \bibinfo{journal}{Phys. Rev.} \textbf{\bibinfo{volume}{D97}},
  \bibinfo{pages}{043525} (\bibinfo{year}{2018}), \eprint{1710.06945}.

\bibitem[{\citenamefont{Nakama et~al.}(2016)\citenamefont{Nakama, Suyama, and
  Yokoyama}}]{Nakama:2016kfq}
\bibinfo{author}{\bibfnamefont{T.}~\bibnamefont{Nakama}},
  \bibinfo{author}{\bibfnamefont{T.}~\bibnamefont{Suyama}}, \bibnamefont{and}
  \bibinfo{author}{\bibfnamefont{J.}~\bibnamefont{Yokoyama}},
  \bibinfo{journal}{Phys. Rev.} \textbf{\bibinfo{volume}{D94}},
  \bibinfo{pages}{103522} (\bibinfo{year}{2016}), \eprint{1609.02245}.

\bibitem[{\citenamefont{Hawking}(1989)}]{Hawking:1987bn}
\bibinfo{author}{\bibfnamefont{S.~W.} \bibnamefont{Hawking}},
  \bibinfo{journal}{Phys. Lett.} \textbf{\bibinfo{volume}{B231}},
  \bibinfo{pages}{237} (\bibinfo{year}{1989}).

\bibitem[{\citenamefont{Polnarev and Zembowicz}(1991)}]{Polnarev:1988dh}
\bibinfo{author}{\bibfnamefont{A.}~\bibnamefont{Polnarev}} \bibnamefont{and}
  \bibinfo{author}{\bibfnamefont{R.}~\bibnamefont{Zembowicz}},
  \bibinfo{journal}{Phys. Rev.} \textbf{\bibinfo{volume}{D43}},
  \bibinfo{pages}{1106} (\bibinfo{year}{1991}).

\bibitem[{\citenamefont{Garriga and Vilenkin}(1993)}]{Garriga:1992nm}
\bibinfo{author}{\bibfnamefont{J.}~\bibnamefont{Garriga}} \bibnamefont{and}
  \bibinfo{author}{\bibfnamefont{A.}~\bibnamefont{Vilenkin}},
  \bibinfo{journal}{Phys. Rev.} \textbf{\bibinfo{volume}{D47}},
  \bibinfo{pages}{3265} (\bibinfo{year}{1993}), \eprint{hep-ph/9208212}.

\bibitem[{\citenamefont{Caldwell and Casper}(1996)}]{Caldwell:1995fu}
\bibinfo{author}{\bibfnamefont{R.~R.} \bibnamefont{Caldwell}} \bibnamefont{and}
  \bibinfo{author}{\bibfnamefont{P.}~\bibnamefont{Casper}},
  \bibinfo{journal}{Phys. Rev.} \textbf{\bibinfo{volume}{D53}},
  \bibinfo{pages}{3002} (\bibinfo{year}{1996}), \eprint{gr-qc/9509012}.

\bibitem[{\citenamefont{Hawking et~al.}(1982)\citenamefont{Hawking, Moss, and
  Stewart}}]{Hawking:1982ga}
\bibinfo{author}{\bibfnamefont{S.~W.} \bibnamefont{Hawking}},
  \bibinfo{author}{\bibfnamefont{I.~G.} \bibnamefont{Moss}}, \bibnamefont{and}
  \bibinfo{author}{\bibfnamefont{J.~M.} \bibnamefont{Stewart}},
  \bibinfo{journal}{Phys. Rev.} \textbf{\bibinfo{volume}{D26}},
  \bibinfo{pages}{2681} (\bibinfo{year}{1982}).

\bibitem[{\citenamefont{Rubin et~al.}(2000)\citenamefont{Rubin, Khlopov, and
  Sakharov}}]{Rubin:2000dq}
\bibinfo{author}{\bibfnamefont{S.~G.} \bibnamefont{Rubin}},
  \bibinfo{author}{\bibfnamefont{M.~{\relax Yu}.} \bibnamefont{Khlopov}},
  \bibnamefont{and} \bibinfo{author}{\bibfnamefont{A.~S.}
  \bibnamefont{Sakharov}}, \bibinfo{journal}{Grav. Cosmol.}
  \textbf{\bibinfo{volume}{6}}, \bibinfo{pages}{51} (\bibinfo{year}{2000}),
  \eprint{hep-ph/0005271}.

\bibitem[{\citenamefont{Khlopov et~al.}(2005)\citenamefont{Khlopov, Rubin, and
  Sakharov}}]{Khlopov:2004sc}
\bibinfo{author}{\bibfnamefont{M.~{\relax Yu}.} \bibnamefont{Khlopov}},
  \bibinfo{author}{\bibfnamefont{S.~G.} \bibnamefont{Rubin}}, \bibnamefont{and}
  \bibinfo{author}{\bibfnamefont{A.~S.} \bibnamefont{Sakharov}},
  \bibinfo{journal}{Astropart. Phys.} \textbf{\bibinfo{volume}{23}},
  \bibinfo{pages}{265} (\bibinfo{year}{2005}), \eprint{astro-ph/0401532}.

\bibitem[{\citenamefont{Garriga et~al.}(2016)\citenamefont{Garriga, Vilenkin,
  and Zhang}}]{Garriga:2015fdk}
\bibinfo{author}{\bibfnamefont{J.}~\bibnamefont{Garriga}},
  \bibinfo{author}{\bibfnamefont{A.}~\bibnamefont{Vilenkin}}, \bibnamefont{and}
  \bibinfo{author}{\bibfnamefont{J.}~\bibnamefont{Zhang}},
  \bibinfo{journal}{JCAP} \textbf{\bibinfo{volume}{1602}}, \bibinfo{pages}{064}
  (\bibinfo{year}{2016}), \eprint{1512.01819}.

\bibitem[{\citenamefont{Deng et~al.}(2017)\citenamefont{Deng, Garriga, and
  Vilenkin}}]{Deng:2016vzb}
\bibinfo{author}{\bibfnamefont{H.}~\bibnamefont{Deng}},
  \bibinfo{author}{\bibfnamefont{J.}~\bibnamefont{Garriga}}, \bibnamefont{and}
  \bibinfo{author}{\bibfnamefont{A.}~\bibnamefont{Vilenkin}},
  \bibinfo{journal}{JCAP} \textbf{\bibinfo{volume}{1704}}, \bibinfo{pages}{050}
  (\bibinfo{year}{2017}), \eprint{1612.03753}.

\bibitem[{\citenamefont{Deng and Vilenkin}(2017)}]{Deng:2017uwc}
\bibinfo{author}{\bibfnamefont{H.}~\bibnamefont{Deng}} \bibnamefont{and}
  \bibinfo{author}{\bibfnamefont{A.}~\bibnamefont{Vilenkin}},
  \bibinfo{journal}{JCAP} \textbf{\bibinfo{volume}{1712}}, \bibinfo{pages}{044}
  (\bibinfo{year}{2017}), \eprint{1710.02865}.

\bibitem[{\citenamefont{Kusenko et~al.}(2020)\citenamefont{Kusenko, Sasaki,
  Sugiyama, Takada, Takhistov, and Vitagliano}}]{Kusenko:2020pcg}
\bibinfo{author}{\bibfnamefont{A.}~\bibnamefont{Kusenko}},
  \bibinfo{author}{\bibfnamefont{M.}~\bibnamefont{Sasaki}},
  \bibinfo{author}{\bibfnamefont{S.}~\bibnamefont{Sugiyama}},
  \bibinfo{author}{\bibfnamefont{M.}~\bibnamefont{Takada}},
  \bibinfo{author}{\bibfnamefont{V.}~\bibnamefont{Takhistov}},
  \bibnamefont{and}
  \bibinfo{author}{\bibfnamefont{E.}~\bibnamefont{Vitagliano}}
  (\bibinfo{year}{2020}), \eprint{2001.09160}.

\bibitem[{\citenamefont{Pani and Loeb}(2013)}]{Pani:2013hpa}
\bibinfo{author}{\bibfnamefont{P.}~\bibnamefont{Pani}} \bibnamefont{and}
  \bibinfo{author}{\bibfnamefont{A.}~\bibnamefont{Loeb}},
  \bibinfo{journal}{Phys. Rev.} \textbf{\bibinfo{volume}{D88}},
  \bibinfo{pages}{041301} (\bibinfo{year}{2013}), \eprint{1307.5176}.

\bibitem[{\citenamefont{Basu et~al.}(1991)\citenamefont{Basu, Guth, and
  Vilenkin}}]{Basu:1991ig}
\bibinfo{author}{\bibfnamefont{R.}~\bibnamefont{Basu}},
  \bibinfo{author}{\bibfnamefont{A.~H.} \bibnamefont{Guth}}, \bibnamefont{and}
  \bibinfo{author}{\bibfnamefont{A.}~\bibnamefont{Vilenkin}},
  \bibinfo{journal}{Phys. Rev.} \textbf{\bibinfo{volume}{D44}},
  \bibinfo{pages}{340} (\bibinfo{year}{1991}).

\bibitem[{\citenamefont{Vilenkin}(1983)}]{Vilenkin:1984hy}
\bibinfo{author}{\bibfnamefont{A.}~\bibnamefont{Vilenkin}},
  \bibinfo{journal}{Phys. Lett.} \textbf{\bibinfo{volume}{133B}},
  \bibinfo{pages}{177} (\bibinfo{year}{1983}).

\bibitem[{\citenamefont{Ipser and Sikivie}(1984)}]{Ipser:1983db}
\bibinfo{author}{\bibfnamefont{J.}~\bibnamefont{Ipser}} \bibnamefont{and}
  \bibinfo{author}{\bibfnamefont{P.}~\bibnamefont{Sikivie}},
  \bibinfo{journal}{Phys. Rev.} \textbf{\bibinfo{volume}{D30}},
  \bibinfo{pages}{712} (\bibinfo{year}{1984}).

\bibitem[{\citenamefont{Carr et~al.}(2017)\citenamefont{Carr, Raidal, Tenkanen,
  Vaskonen, and Veerm{\"a}e}}]{Carr:2017jsz}
\bibinfo{author}{\bibfnamefont{B.}~\bibnamefont{Carr}},
  \bibinfo{author}{\bibfnamefont{M.}~\bibnamefont{Raidal}},
  \bibinfo{author}{\bibfnamefont{T.}~\bibnamefont{Tenkanen}},
  \bibinfo{author}{\bibfnamefont{V.}~\bibnamefont{Vaskonen}}, \bibnamefont{and}
  \bibinfo{author}{\bibfnamefont{H.}~\bibnamefont{Veerm{\"a}e}},
  \bibinfo{journal}{Phys. Rev.} \textbf{\bibinfo{volume}{D96}},
  \bibinfo{pages}{023514} (\bibinfo{year}{2017}), \eprint{1705.05567}.

\bibitem[{\citenamefont{Ricotti et~al.}(2008)\citenamefont{Ricotti, Ostriker,
  and Mack}}]{Ricotti:2007au}
\bibinfo{author}{\bibfnamefont{M.}~\bibnamefont{Ricotti}},
  \bibinfo{author}{\bibfnamefont{J.~P.} \bibnamefont{Ostriker}},
  \bibnamefont{and} \bibinfo{author}{\bibfnamefont{K.~J.} \bibnamefont{Mack}},
  \bibinfo{journal}{Astrophys. J.} \textbf{\bibinfo{volume}{680}},
  \bibinfo{pages}{829} (\bibinfo{year}{2008}), \eprint{0709.0524}.

\bibitem[{\citenamefont{Ali-Ha{\"\i}moud and
  Kamionkowski}(2017)}]{Ali-Haimoud:2016mbv}
\bibinfo{author}{\bibfnamefont{Y.}~\bibnamefont{Ali-Ha{\"\i}moud}}
  \bibnamefont{and}
  \bibinfo{author}{\bibfnamefont{M.}~\bibnamefont{Kamionkowski}},
  \bibinfo{journal}{Phys. Rev.} \textbf{\bibinfo{volume}{D95}},
  \bibinfo{pages}{043534} (\bibinfo{year}{2017}), \eprint{1612.05644}.

\bibitem[{\citenamefont{Weinberg}(1971)}]{Weinberg:1971mx}
\bibinfo{author}{\bibfnamefont{S.}~\bibnamefont{Weinberg}},
  \bibinfo{journal}{Astrophys. J.} \textbf{\bibinfo{volume}{168}},
  \bibinfo{pages}{175} (\bibinfo{year}{1971}).

\bibitem[{\citenamefont{Deng et~al.}(2018)\citenamefont{Deng, Vilenkin, and
  Yamada}}]{Deng:2018cxb}
\bibinfo{author}{\bibfnamefont{H.}~\bibnamefont{Deng}},
  \bibinfo{author}{\bibfnamefont{A.}~\bibnamefont{Vilenkin}}, \bibnamefont{and}
  \bibinfo{author}{\bibfnamefont{M.}~\bibnamefont{Yamada}},
  \bibinfo{journal}{JCAP} \textbf{\bibinfo{volume}{1807}}, \bibinfo{pages}{059}
  (\bibinfo{year}{2018}), \eprint{1804.10059}.

\bibitem[{\citenamefont{Khatri and Sunyaev}(2015)}]{Khatri:2015tla}
\bibinfo{author}{\bibfnamefont{R.}~\bibnamefont{Khatri}} \bibnamefont{and}
  \bibinfo{author}{\bibfnamefont{R.}~\bibnamefont{Sunyaev}},
  \bibinfo{journal}{JCAP} \textbf{\bibinfo{volume}{1509}}, \bibinfo{pages}{026}
  (\bibinfo{year}{2015}), \eprint{1507.05615}.

\bibitem[{\citenamefont{Sunyaev and Zeldovich}(1970)}]{Sunyaev:1970er}
\bibinfo{author}{\bibfnamefont{R.~A.} \bibnamefont{Sunyaev}} \bibnamefont{and}
  \bibinfo{author}{\bibfnamefont{{\relax Ya}.~B.} \bibnamefont{Zeldovich}},
  \bibinfo{journal}{Astrophys. Space Sci.} \textbf{\bibinfo{volume}{7}},
  \bibinfo{pages}{20} (\bibinfo{year}{1970}).

\bibitem[{\citenamefont{Daly}(1991)}]{Daly}
\bibinfo{author}{\bibfnamefont{R.}~\bibnamefont{Daly}}, \bibinfo{journal}{The
  Astrophysical Journal} \textbf{\bibinfo{volume}{371}}, \bibinfo{pages}{14}
  (\bibinfo{year}{1991}).

\bibitem[{\citenamefont{Hu and Silk}(1993)}]{Hu:1992dc}
\bibinfo{author}{\bibfnamefont{W.}~\bibnamefont{Hu}} \bibnamefont{and}
  \bibinfo{author}{\bibfnamefont{J.}~\bibnamefont{Silk}},
  \bibinfo{journal}{Phys. Rev.} \textbf{\bibinfo{volume}{D48}},
  \bibinfo{pages}{485} (\bibinfo{year}{1993}).

\bibitem[{\citenamefont{Chluba et~al.}(2012)\citenamefont{Chluba, Erickcek, and
  Ben-Dayan}}]{Chluba:2012we}
\bibinfo{author}{\bibfnamefont{J.}~\bibnamefont{Chluba}},
  \bibinfo{author}{\bibfnamefont{A.~L.} \bibnamefont{Erickcek}},
  \bibnamefont{and}
  \bibinfo{author}{\bibfnamefont{I.}~\bibnamefont{Ben-Dayan}},
  \bibinfo{journal}{Astrophys. J.} \textbf{\bibinfo{volume}{758}},
  \bibinfo{pages}{76} (\bibinfo{year}{2012}), \eprint{1203.2681}.

\bibitem[{\citenamefont{Chluba}(2016)}]{Chluba:2016bvg}
\bibinfo{author}{\bibfnamefont{J.}~\bibnamefont{Chluba}},
  \bibinfo{journal}{Mon. Not. Roy. Astron. Soc.}
  \textbf{\bibinfo{volume}{460}}, \bibinfo{pages}{227} (\bibinfo{year}{2016}),
  \eprint{1603.02496}.

\bibitem[{\citenamefont{Fixsen et~al.}(1996)\citenamefont{Fixsen, Cheng, Gales,
  Mather, Shafer, and Wright}}]{Fixsen:1996nj}
\bibinfo{author}{\bibfnamefont{D.~J.} \bibnamefont{Fixsen}},
  \bibinfo{author}{\bibfnamefont{E.~S.} \bibnamefont{Cheng}},
  \bibinfo{author}{\bibfnamefont{J.~M.} \bibnamefont{Gales}},
  \bibinfo{author}{\bibfnamefont{J.~C.} \bibnamefont{Mather}},
  \bibinfo{author}{\bibfnamefont{R.~A.} \bibnamefont{Shafer}},
  \bibnamefont{and} \bibinfo{author}{\bibfnamefont{E.~L.}
  \bibnamefont{Wright}}, \bibinfo{journal}{Astrophys. J.}
  \textbf{\bibinfo{volume}{473}}, \bibinfo{pages}{576} (\bibinfo{year}{1996}),
  \eprint{astro-ph/9605054}.

\bibitem[{\citenamefont{Hill et~al.}(2015)\citenamefont{Hill, Battaglia,
  Chluba, Ferraro, Schaan, and Spergel}}]{Hill:2015tqa}
\bibinfo{author}{\bibfnamefont{J.~C.} \bibnamefont{Hill}},
  \bibinfo{author}{\bibfnamefont{N.}~\bibnamefont{Battaglia}},
  \bibinfo{author}{\bibfnamefont{J.}~\bibnamefont{Chluba}},
  \bibinfo{author}{\bibfnamefont{S.}~\bibnamefont{Ferraro}},
  \bibinfo{author}{\bibfnamefont{E.}~\bibnamefont{Schaan}}, \bibnamefont{and}
  \bibinfo{author}{\bibfnamefont{D.~N.} \bibnamefont{Spergel}},
  \bibinfo{journal}{Phys. Rev. Lett.} \textbf{\bibinfo{volume}{115}},
  \bibinfo{pages}{261301} (\bibinfo{year}{2015}), \eprint{1507.01583}.

\bibitem[{\citenamefont{Serpico et~al.}(2020)\citenamefont{Serpico, Poulin,
  Inman, and Kohri}}]{Serpico:2020ehh}
\bibinfo{author}{\bibfnamefont{P.~D.} \bibnamefont{Serpico}},
  \bibinfo{author}{\bibfnamefont{V.}~\bibnamefont{Poulin}},
  \bibinfo{author}{\bibfnamefont{D.}~\bibnamefont{Inman}}, \bibnamefont{and}
  \bibinfo{author}{\bibfnamefont{K.}~\bibnamefont{Kohri}}
  (\bibinfo{year}{2020}), \eprint{2002.10771}.

\end{thebibliography}
\end{document}